\documentclass[aps,prl,groupedaddress,twocolumn,showpacs,floatfix
]{revtex4-1}  
\usepackage{txfonts}  
\usepackage{graphicx,bm}  

\def\be{\begin{equation}}
\def\ee{\end{equation}}
\def\ba{\begin{eqnarray}}
\def\ea{\end{eqnarray}}
\def\bc{\begin{center}}
\def\ec{\end{center}}

\begin{document}  

\title{Nonlinear electromagnetic response of a uniform electron gas}  

\author{S. A. Mikhailov}  
\email[Electronic mail: ]{sergey.mikhailov@physik.uni-augsburg.de}  

\affiliation{Institute of Physics, University of Augsburg, D-86135 Augsburg, Germany}  

\date{\today}  

\begin{abstract}  
The linear electromagnetic response of a uniform electron gas to a longitudinal electric field is determined, within the self-consistent-field theory, by the linear polarizability and the Lindhard dielectric function. Using the same approach we derive  analytical expressions for the second- and third-order nonlinear polarizabilities of the three-, two- and one-dimensional homogeneous electron gases with the parabolic electron energy dispersion. The results are valid both for degenerate (Fermi) and non-degenerate (Boltzmann) electron gases. A resonant enhancement of the second and third harmonics generation due to a combination of the single-particle and collective (plasma) resonances is predicted. 
\end{abstract}  

\pacs{78.67.-n, 
71.10.Ca, 
73.20.Mf, 
42.65.Ky
}

\maketitle  

The interaction of the electromagnetic radiation with a gaseous or solid-state plasma is well described, within the linear-response theory, by the Lindhard dielectric function $\epsilon(\bm q,\omega)$ \cite{Lindhard54}. Being originally derived for a three-dimensional (3D) uniform electron gas \cite{Lindhard54}, it was generalized to the two- (2D) and one-dimensional (1D) electron systems by Stern \cite{Stern67} and Das Sarma with coauthors \cite{Li89}. In this theory electron-electron interaction is taken into account within the self-consistent mean-field approach which is equivalent \cite{Ehrenreich59} to the random phase approximation (RPA). This theory was shown to be very accurate in describing the electromagnetic response and plasma oscillations of the uniform electron gas \cite{Mahan90}, both in 3D, and in lower dimensions -- in semiconductor quantum-well, -wire, and -dot structures, see, e.g.,  \cite{Allen77,Demel88,Kukushkin05}.

The {\em nonlinear} electromagnetic response of a uniform electron gas in low-dimensional systems is studied in much less detail. A possible reason for that consisted in the low quality of solid-state structures: while in the gaseous plasma collisions of electrons and ions do not play a significant role, which allows one to observe the nonlinear phenomena in relatively low external electric fields, in solids the scattering and disorder effects were quite strong which required very large  electric fields and hindered the observation of the nonlinear phenomena. 

The progress of semiconductor technology changed this situation in recent years. It has become possible to create semiconductor GaAs/AlGaAs quantum-well structures with the electron mobility $\mu\simeq 3\times 10^7$ cm$^2$/Vs \cite{Pfeiffer03} which corresponds to the electron mean-free-path comparable with the sample dimensions ($l_{mfp}\simeq 1-2$ mm). In such systems a strongly nonlinear electrodynamic effect -- the giant microwave induced magnetoresistance oscillations manifesting themselves in relatively low ac electric fields ($\lesssim 1$ V/cm) -- was recently discovered \cite{Mani02,Zudov03} and attracted much attention (for an overview of the state of the art and further references see \cite{Mikhailov14a}). The experiments \cite{Mani02,Zudov03} have been explained \cite{Mikhailov11a,Mikhailov14a} by the influence of ponderomotive forces, which are usually very small in the fields $\simeq 1$ V/cm but become sufficiently strong in the ultra clean samples \cite{Pfeiffer03,Mani02,Zudov03}, especially near internal electron resonances (near the cyclotron resonance harmonics in Refs. \cite{Mani02,Zudov03}). In Ref. \cite{Mikhailov14a} it was also shown that, under the same experimental conditions not only the nonlinear transport phenomena (the microwave photoconductivity \cite{Mani02,Zudov03}) but also the high-frequency nonlinear effects (harmonics generation, frequency mixing) should be observed in such a collisionless 2D electron plasma. Another nonlinear electrodynamic effect -- the giant enhancement of the second harmonic -- was predicted in the clean 2D electron systems  near the plasma resonance in zero magnetic field \cite{Mikhailov11c}. The availability of the almost collisionless 2D electron plasma in GaAs quantum-well structures thus offers great opportunities to study nonlinear electrodynamic phenomena in the easily achievable ac electric fields. A theoretical study of such nonlinear effects thus becomes highly topical and very desirable.

In this Letter we generalize the Lindhard linear-response theory to the case of the nonlinear response. We derive closed-form analytical expressions for the second- and third-order nonlinear polarizabilities of a uniform electron gas in zero magnetic field. Our results are valid for 3D, 2D and 1D electron gases, both for degenerate (Fermi) and non-degenerate (Boltzmann) statistics. The second order response function, similar to $\epsilon(\bm q,\omega)$, is derived for all values of the wave-vector $\bm q$ and the frequency $\omega$. The optimal conditions of the second and third harmonic generation strongly enhanced by the single-particle and collective (plasma) resonances are derived and discussed. 

We consider a $d$-dimensional ($d=1,2,3$) uniform electron gas under the action of the electric field ${\bm E}({\bm r},t)=-\nabla \phi({\bm r},t)$ described by the potential
\be 
\phi({\bm r},t)=\phi_{{\bm q}\omega}e^{i{\bm q}\cdot {\bm r}-i(\omega+i\gamma) t}    + \textrm{c.c.} ,\label{input-potential}
\ee
where $\gamma\to +0$, ${\bm q}$-vector is $d$-dimensional, and c.c. means the complex conjugate. The linear response of such a system is generally described by two functions: the relation between the potential $\phi_{{\bm q}\omega}$ and the induced charge density fluctuation $\rho_{{\bm q}\omega}$, 
\be 
\rho_{{\bm q}\omega}
=\alpha^{(1),d}_{{\bm q}\omega;{\bm q}\omega}({\bm q},\omega)
\phi_{{\bm q}\omega}, \label{rho-phi}
\ee
and the relation between the external $\phi_{{\bm q}\omega}^{\rm ext}$, induced $\phi_{{\bm q}\omega}^{\rm ind}$, and total $\phi_{{\bm q}\omega}\equiv \phi_{{\bm q}\omega}^{\rm tot}$ potentials, 
\be 
\phi_{{\bm q}\omega}^{\rm tot}=\phi_{{\bm q}\omega}^{\rm ext}+\phi_{{\bm q}\omega}^{\rm ind}={\cal R}^{(1),d}_{{\bf q}\omega;{\bf q}\omega}({\bf q},\omega)\phi_{{\bm q}\omega}^{\rm ext}.
\ee
The function $\alpha^{(1),d}_{{\bm q}\omega;{\bm q}\omega}$ is the first-order polarizability of the $d$-dimensional electron gas. The function ${\cal R}^{(1),d}_{{\bf q}\omega;{\bf q}\omega}$ takes into account Coulomb interaction between electrons within the self-consistent-field approach \cite{Stern67,Li89,Ehrenreich59} and is related to the dielectric function $\epsilon_d(\bm q,\omega)$,
\be 
{\cal R}^{(1),d}_{{\bf q}\omega;{\bf q}\omega}({\bf q},\omega)=\frac{1}{\epsilon_d(\bm q,\omega)}\equiv\frac 1{1-V^C_d(q)\alpha^{(1),d}_{{\bf q}\omega;{\bf q}\omega}({\bf q},\omega)};
\label{R1}
\ee
here  $V^C_d(q)$ is the Fourier transform of the Coulomb potential in $d$ dimensions [in 3D and 2D $V^C_3(q)=4\pi/q^2$ and $V^C_2(q)=2\pi/q$, respectively]. Poles of the response function ${\cal R}^{(1),d}_{{\bf q}\omega;{\bf q}\omega}({\bf q},\omega)$, i.e. zeros of the dielectric function $\epsilon_d(\bm q,\omega)$, determine the spectrum of plasma waves in the system.

To describe the nonlinear response of the uniform $d$-dimensional electron gas we introduce the functions
\ba 
\rho_{2{\bm q}2\omega}&=&
\alpha^{(2),d}_{2{\bm q}2\omega;{\bm q}\omega,{\bm q}\omega}({\bm q},\omega)(\phi_{{\bm q}\omega})^2,
\nonumber \\ 
 \rho_{3{\bm q}3\omega}
&=&
\alpha^{(3),d}_{3{\bm q}3\omega;{\bm q}\omega,{\bm q}\omega,{\bm q}\omega}({\bm q},\omega)
(\phi_{{\bm q}\omega})^3, \label{polariz-definitions}
\\
\phi^{\rm tot}_{2{\bm q}2\omega}&=&{\cal R}^{(2),d}_{2{\bm q}2\omega;{\bm q}\omega{\bm q}\omega}({\bm q},\omega)\left(\phi^{\rm ext}_{{\bm q}\omega}\right)^2.
 \nonumber
\ea
The first superscript here (in parenthesis) stands for the order of the response; the second superscript $d$ ($d=1,2$ or $3$) is the dimensionality of the electron gas. The subscripts designate the specific response process; for example, the third-order function $\alpha^{(3),d}_{3{\bm q}3\omega;{\bm q}\omega,{\bm q}\omega,{\bm q}\omega}$ describes the process ``three quanta ${\bm q}\omega$ come in, one quantum $3{\bm q}3\omega$ goes out'' [we consider only the harmonic generation effects ignoring other nonlinear processes like, e.g., $\{({\bm q}\omega),-({\bm q}\omega),({\bm q}\omega)\}\to ({\bm q}\omega$)]. 

The nine quantities $\alpha^{(o),d}$ ($o=1,2,3$, $d=1,2,3$) are measured in different units. To calculate them and to present results in a universal form it is convenient to introduce dimensionless quantities $\pi^{(o),d}$ related to $\alpha^{(o),d}$ as follows:
\ba 
\pi^{(1),d}_{{\bm q}\omega;{\bm q}\omega}({\bm q},\omega)&=&a_B^{d-1} \alpha^{(1),d}_{{\bm q}\omega;{\bm q}\omega}({\bm q},\omega),
\nonumber \\ 
\pi^{(2),d}_{2{\bm q}2\omega;{\bm q}\omega,{\bm q}\omega}({\bm q},\omega)&=&
-ea_B^{d-2}\alpha^{(2),d}_{2{\bm q}2\omega;{\bm q}\omega,{\bm q}\omega}({\bm q},\omega),
\\ 
\pi^{(3),d}_{3{\bm q}3\omega;{\bm q}\omega,{\bm q}\omega,{\bm q}\omega}({\bm q},\omega)&=&
e^2a_B^{d-3}\alpha^{(3),d}_{3{\bm q}3\omega;{\bm q}\omega,{\bm q}\omega,{\bm q}\omega}({\bm q},\omega);\nonumber 
\ea
then the charge densities $\rho_{{\bm q}\omega}$, $\rho_{2{\bm q}2\omega}$ and $\rho_{3{\bm q}3\omega}$ in the $d$-dimensional electron gas are measured in units $-e/a_B^d$ and the potential $\phi_{{\bm q}\omega}$ -- in units $-e/a_B$, where $a_B$ is the effective Bohr radius and $e>0$ is the electron charge. Solving the quantum kinetic equation for the density matrix one gets the following expressions for the first- and second-order polarizabilities 
\begin{widetext}
\be 
\pi^{(1),d}_{{\bm q}\omega;{\bm q}\omega}
=\frac {e^2a_B^{d-1}g_s}{L^d}
\sum_{\bm k \bm k'} 
\frac{f_0(E_{\bm k'})-f_0(E_{\bm k})} {E_{\bm k'}-E_{\bm k}+\hbar\omega+i\hbar\gamma}
 \langle {\bm k}'|e^{-i{\bm q}\cdot {\bm r}}|{\bm k}\rangle
\langle{\bm k}|e^{i{\bm q}\cdot {\bm r}} |{\bm k}'\rangle ,
\label{pi1}
\ee
\be
\pi^{(2),d}_{2{\bm q}2\omega;{\bm q}\omega,{\bm q}\omega}=
\frac {e^4a_B^{d-2}g_s}{L^d}
\sum_{\bm{kk}'\bm{k}''} 
\frac{
\langle \bm{k}' |e^{-i2{\bm q}\cdot {\bm r}}|\bm{k}\rangle
\langle\bm{k}|e^{i{\bm q}\cdot {\bm r}} |\bm{k}''\rangle
\langle\bm{k}''|  e^{i{\bm q}\cdot {\bm r}} |\bm{k}'\rangle
} {E_{\bm{k}'}-E_{\bm{k}}+2\hbar\omega +2i\hbar\gamma}
\Bigg\{
\frac{f_0(E_{\bm{k}'})-f_0(E_{\bm{k}''})} {E_{\bm{k}'}-E_{\bm{k}''}+\hbar\omega+i\hbar\gamma}
-
\frac{f_0(E_{\bm{k}''})-f_0(E_{\bm{k}})} {E_{\bm{k}''}-E_{\bm{k}}+\hbar\omega+i\hbar\gamma}
\Bigg\},
\label{pi2}
\ee
\end{widetext}
and a similar expression for $\pi^{(3),d}_{3{\bm q}3\omega;{\bm q}\omega,{\bm q}\omega,{\bm q}\omega}$ which we do not present here for brevity; here $g_s=2$ is the spin degeneracy factor, $|{\bm k}\rangle=e^{i{\bm k\cdot \bm r}}/L^{d/2}$ is the electron wave function, $E_{\bm k}=\hbar^2k^2/2m^\star$ (we consider electrons with the parabolic energy dispersion thus excluding the case of graphene, cf. Ref. \cite{Mikhailov11c}), $f_0(E)=\left\{1+\exp[(E-\mu)/T]\right\}^{-1}$ is the equilibrium Fermi-Dirac distribution function, $\mu$ is the chemical potential and $T$ is the temperature. If the electron gas is degenerate and $T=0$, the functions $\pi^{(1),d}_{{\bm q}\omega;{\bm q}\omega}$, Eq. (\ref{pi1}), can be analytically calculated for any $d$, see Refs.  \cite{Lindhard54,Stern67,Li89}; for example, for the 2D electron gas one gets \cite{Stern67}: 
\be
\pi^{(1),2}_{{\bm q}\omega;{\bm q}\omega}({\bf q},\omega)=
\frac {g_s }{2\pi Q}
\left\{
F_2\left(\frac{\Omega}{2Q}-\frac Q2\right)
-F_2\left(\frac{\Omega}{2Q}+\frac Q2\right)
\right\},
\label{pi2d}
\ee
where $Q= q/k_F$, $\Omega=\hbar\omega/E_F$, $E_F$ and $k_F$ are the Fermi energy and momentum, and 
\be 
F_2(x)=
\left\{
\begin{array}{lc}
  x + \sqrt{x^2-1}  , & x<-1 \\
  x -i\sqrt{1- x^2} , & -1<x<1 \\
  x -\sqrt{x^2-1}   , & 1<x \\
\end{array}
\right.
\label{F2}.
\ee
In order to find the second- and third-order polarizabilities one should calculate integrals in Eq. (\ref{pi2}) and in the corresponding expression for $\pi^{(3),d}_{3{\bm q}3\omega;{\bm q}\omega,{\bm q}\omega,{\bm q}\omega}
$. Although these formulas look rather cumbersome, one can straightforwardly show that 
\begin{widetext}
\be
\pi^{(2),d}_{2{\bm q}2\omega;{\bm q}\omega,{\bm q}\omega}({\bm q},\omega)=
2\frac {
\pi^{(1),d}_{{\bm q}\omega;{\bm q}\omega}(2{\bm q},2\omega)
-
\pi^{(1),d}_{{\bm q}\omega;{\bm q}\omega}({\bm q},\omega)}{q^2a_B^2},
\label{relationPi2Pi1}
\ee
\be 
\pi^{(3),d}_{3{\bm q}3\omega;{\bm q}\omega,{\bm q}\omega,{\bm q}\omega}({\bm q},\omega)
=\frac {3\pi^{(1),d}_{{\bm q}\omega;{\bm q}\omega}(3{\bm q},3\omega)
-8\pi^{(1),d}_{{\bm q}\omega;{\bm q}\omega}(2{\bm q},2\omega)
+5\pi^{(1),d}_{{\bm q}\omega;{\bm q}\omega}({\bm q},\omega)}{ 3(qa_B)^4} \label{chi(3)}.
\ee
\end{widetext}
The relations (\ref{relationPi2Pi1}) -- (\ref{chi(3)}) are valid for any dimensionality of the electron gas ($d=1,2,3$) and for any relation between the chemical potential $\mu$ and the temperature $T$, i.e. both for degenerate (Fermi) and non-degenerate (Boltzmann) electron gases. If the electron gas is degenerate ($T=0$), Eqs. (\ref{relationPi2Pi1}) -- (\ref{chi(3)}) provide, together with the corresponding linear-response results \cite{Lindhard54,Stern67,Li89}, analytical expressions for the second- and third-order polarizabilities. 

\begin{figure}
\includegraphics[width=8.4cm]{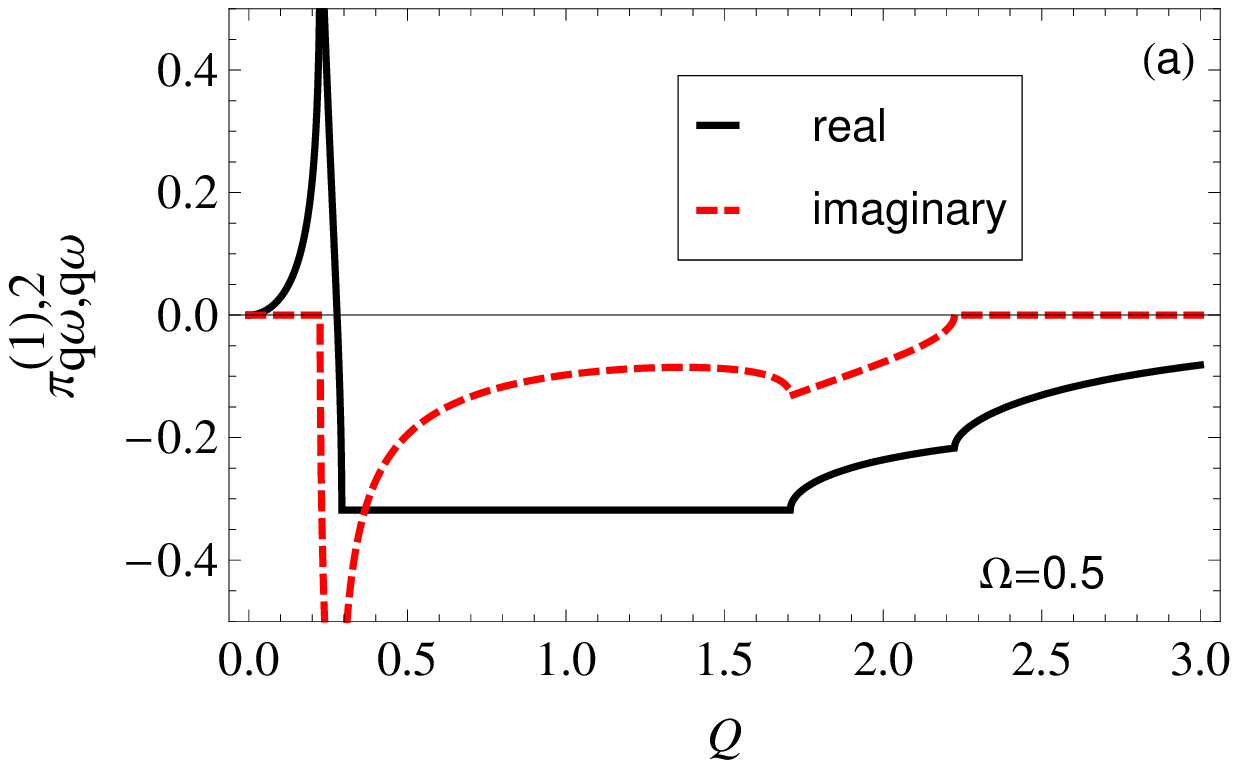}
\includegraphics[width=8.4cm]{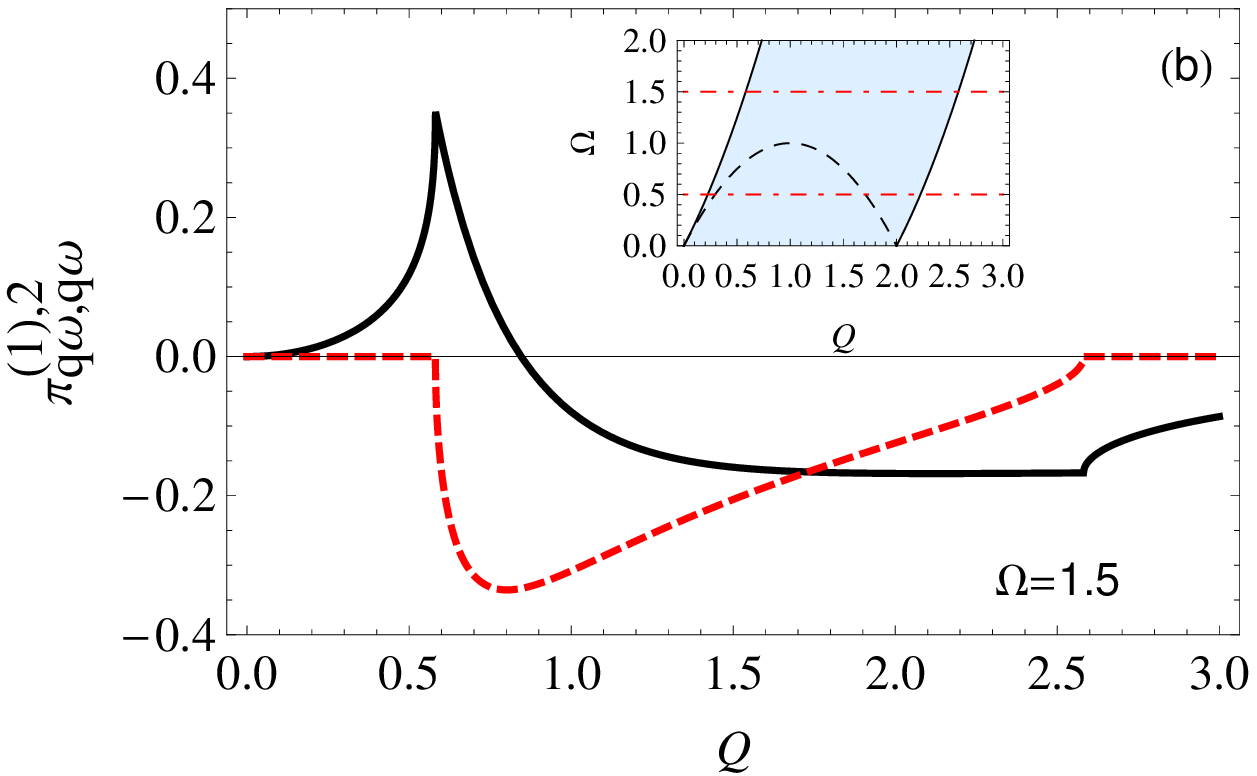}
\caption{\label{fig:chi12} The real (black solid curves) and imaginary (red dashed curves) parts of the first-order polarizability of the 2D electron gas (\ref{pi2d}) as a function of the wave-vector $Q$ at (a) $\Omega=0.5$ and (b) $\Omega=1.5$, Ref. \cite{Stern67}. The inset in (b) shows the single-particle absorption continuum (shaded area) bounded by the curves (\ref{sp-bound}) with $o=1$. The red dash-dotted lines in the Inset show the values of $\Omega$ for which the polarizability $\pi^{(1),2}_{{\bm q}\omega;{\bm q}\omega}$ is plotted on the main plots.}
\end{figure}

Figure \ref{fig:chi12} shows the first-order polarizability of the 2D electron gas (\ref{pi2d}), Ref. \cite{Stern67}, as a function of the wave-vector at two different values of the frequency. The points of the discontinuous derivatives are related to the boundaries of the single-particle absorption areas shown in the Inset to Figure \ref{fig:chi12}(b) and determined by the curves
\be 
\Omega= 2Q+oQ^2,\ \ \Omega= 2Q-oQ^2,\ \ \Omega= -2Q+oQ^2\label{sp-bound}
\ee
with $o=1$. One sees that at small values of $Q$ the imaginary part of $\pi^{(1),2}_{{\bm q}\omega;{\bm q}\omega}$ vanishes and its real part has a sharp maximum when $Q$ touches the left boundary of the single-particle absorption area $\Omega= 2Q+Q^2$.

\begin{figure}
\includegraphics[width=8.4cm]{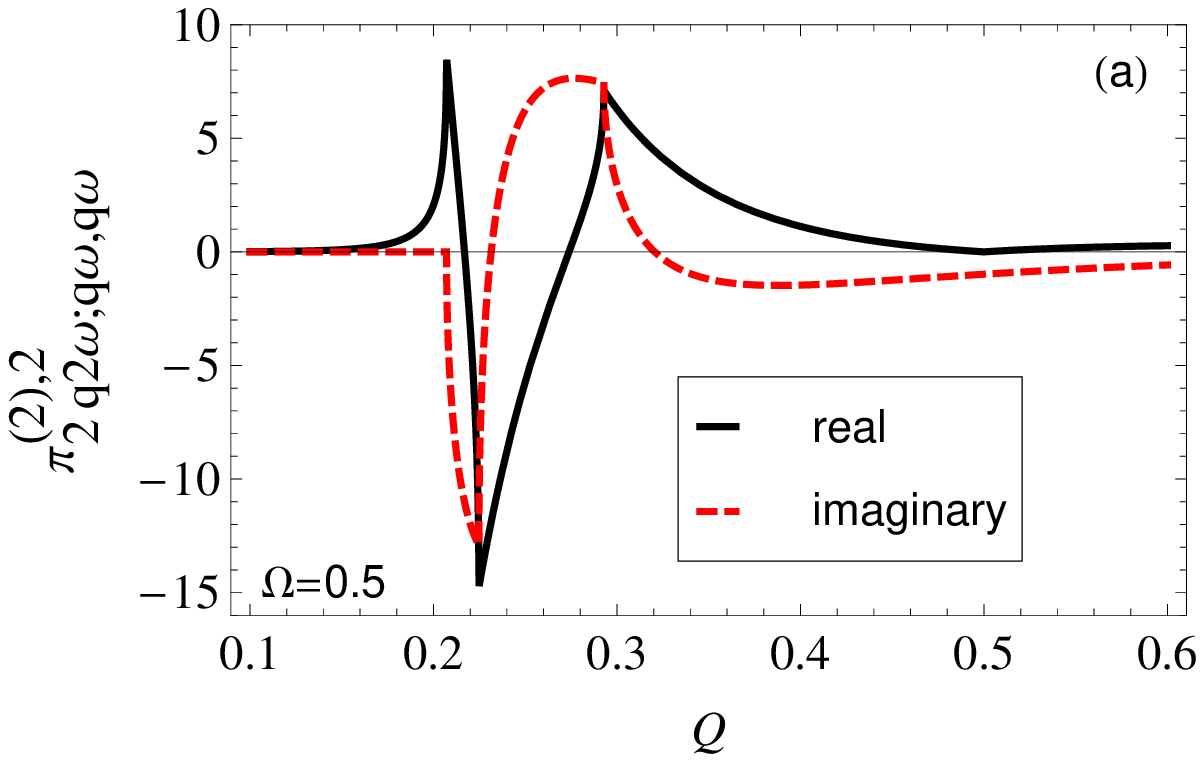}
\includegraphics[width=8.4cm]{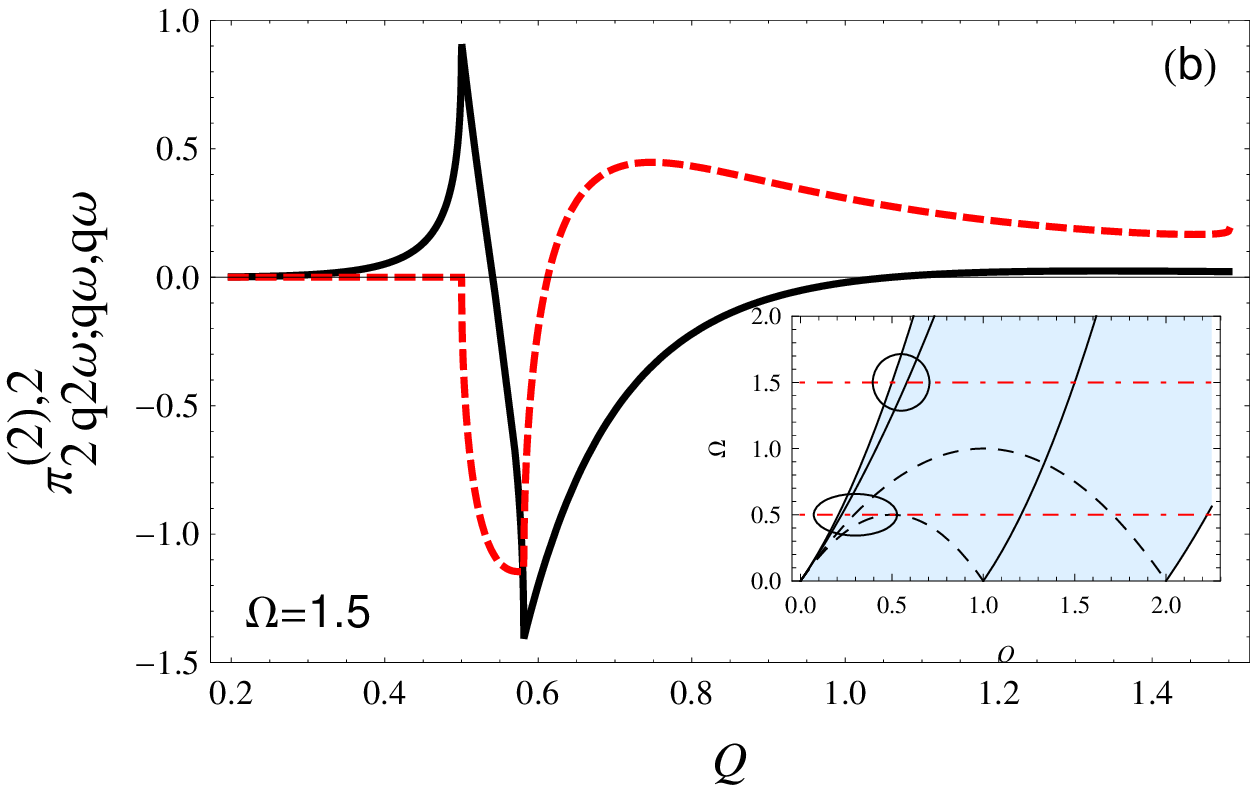}
\caption{\label{fig:chi22} The second-order polarizability of the 2D electron gas (\ref{relationPi2Pi1}) at $k_Fa_B=1$. The inset in (b) shows the single-particle absorption (shaded) areas, bounded by the curves (\ref{sp-bound}) with $o=1$ and $o=2$. The range of $Q$ shown in (a) and (b) is encircled in the Inset. 
}
\end{figure}

\begin{figure}
\includegraphics[width=8.4cm]{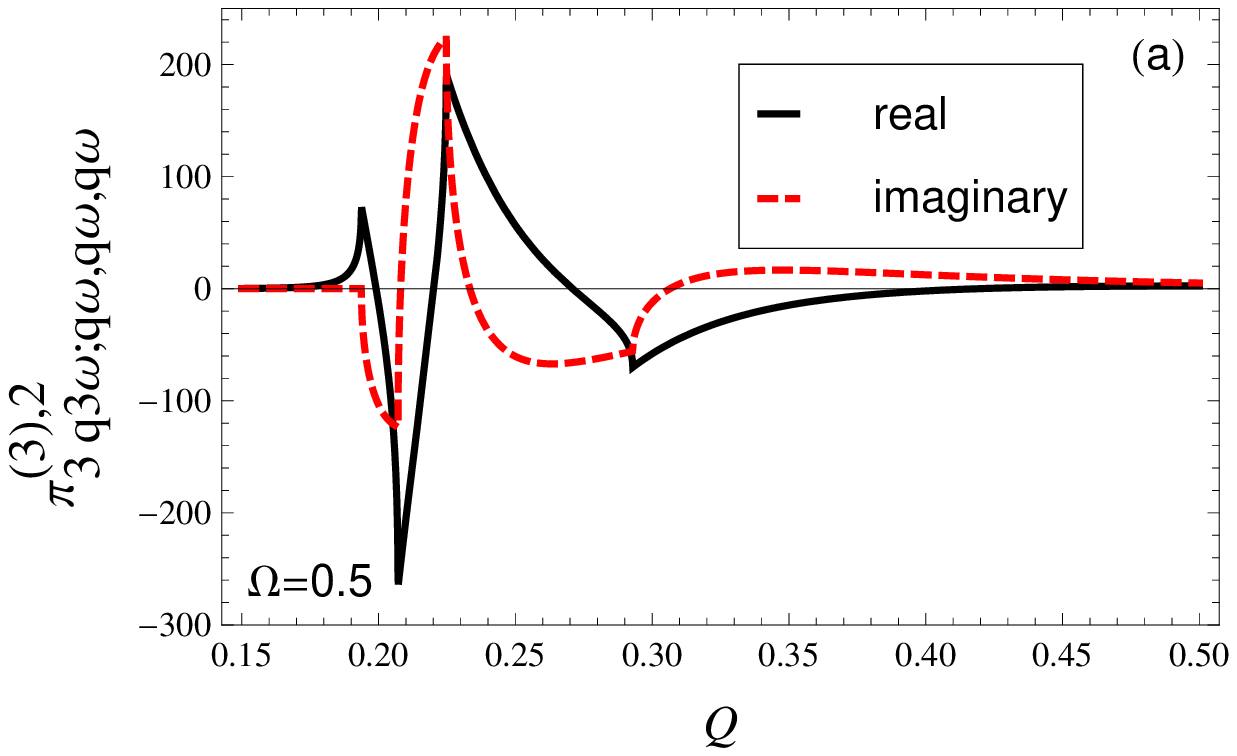}
\includegraphics[width=8.4cm]{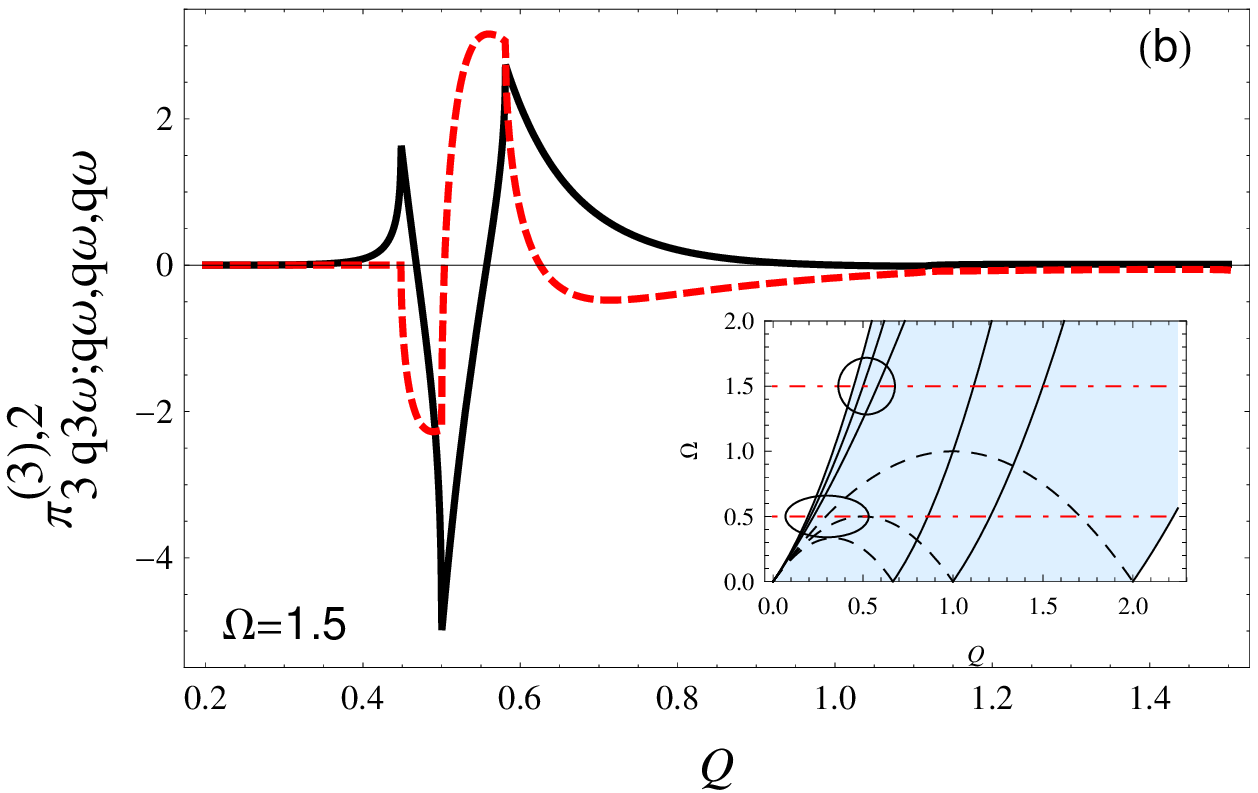}
\caption{\label{fig:chi32} The same as in Figures \ref{fig:chi12} and \ref{fig:chi22} but for the third-order polarizability of the 2D electron gas (\ref{chi(3)}) at $k_Fa_B=1$. 
}
\end{figure}

Figure \ref{fig:chi22} shows the second-order polarizability of the 2D electron gas (\ref{relationPi2Pi1}) as a function of $Q$ at the same values of the frequency $\Omega$. Due to the function $\pi^{(1),d}_{{\bm q}\omega;{\bm q}\omega}(2{\bm q},2\omega)$ in the right-hand side of (\ref{relationPi2Pi1}) the single-particle absorption area consists now of two overlapping areas [see Inset to Figure \ref{fig:chi22}(b)], bounded by the curves (\ref{sp-bound}) with $o=1$ and $o=2$. The same calculation can be done for the third-order polarizability, see Figure \ref{fig:chi32}. All three polarizabilities $\pi^{(o),d}$ have a sharp maximum when the point $(Q,\Omega)$ approaches the left boundary of the single-particle absorption area. At small $Q$ the maxima of $\pi^{(3),d}$ are larger than those of $\pi^{(2),d}$ and of $\pi^{(1),d}$, compare Figs. \ref{fig:chi32}(a) with \ref{fig:chi22}(a) and \ref{fig:chi12}(a). 

\begin{figure}
\includegraphics[width=8cm]{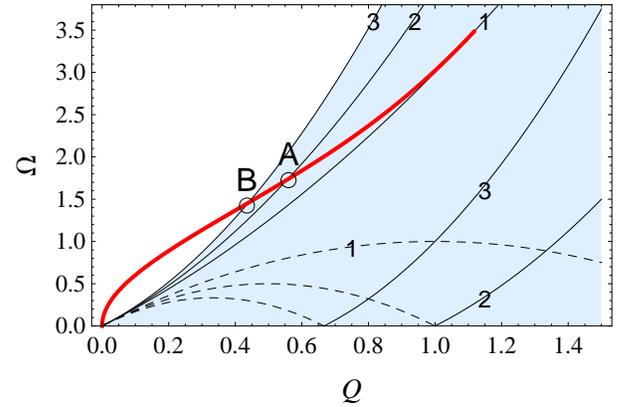}
\caption{\label{fig:plasmon} The spectrum of 2D plasmons (thick red curve) at $k_Fa_B=1.2$. The numbers 1, 2, 3 on the curves mark the single-particle absorption areas (\ref{sp-bound}) with $o=1,2,3$. The maximum of the second- (third-)order response function is achieved at the point A (B) corresponding to the intersection of the 2D plasmon curve with the second- and third-order boundary curves (\ref{sp-bound}). 
}
\end{figure}

Now consider the second-order self-consistent response function ${\cal R}^{(2),d}_{2{\bm q}2\omega;{\bm q}\omega{\bm q}\omega}({\bm q},\omega)$. For a 2D electron gas it was derived in Ref. \cite{Mikhailov11c}, see also \cite{Mikhailov12a}. In the general case of any dimensionality it assumes the form
\be 
{\cal R}^{(2),d}_{2{\bm q}2\omega;{\bm q}\omega{\bm q}\omega}({\bm q},\omega)= V^C_d(2q)
\frac{\alpha_{2{\bm q}2\omega;{\bm q}\omega,{\bm q}\omega}^{(2),d}({\bm q},\omega)
}{\epsilon_d(2{\bm q},2\omega)\left[\epsilon_d({\bm q},\omega)\right]^2}.
\label{R2}
\ee
This function determines, according to (\ref{polariz-definitions}), the total potential in the system with the double wave-vector $2\bm q$ and the double frequency $2\omega$ (the second harmonic intensity). This function is proportional to the second order polarizability $\alpha_{2{\bm q}2\omega;{\bf q}\omega,{\bm q}\omega}^{(2),d}$ and has a sharp maximum when $\Omega=2Q+2Q^2$ (the single-particle resonance). In addition, the function (\ref{R2}) has two poles at the 2D plasmon frequencies corresponding to the zeros of the dielectric function $\epsilon_d({\bf q},\omega)$ (the second-order pole) and $\epsilon_d(2{\bm q},2\omega)$ (the first-order pole). Near the points of the $Q$-$\Omega$ plane where the single-particle resonance coincides with the collective (plasmon) resonance one should expect a giant growth of the second harmonic intensity. A similar effect is expected at the third harmonic. Figure \ref{fig:plasmon} illustrates the best conditions for the second and third harmonic generation in a 2D electron system. Similar (and even stronger) effects should be also seen in a 1D structures (quantum wires).

The predicted effects can be experimentally observed in the high-electron mobility GaAs/AlGaAs quantum wells or wires in standard geometries used for the excitation of the 2D (1D) plasmons \cite{Allen77,Demel88,Kukushkin05} (with a grating coupler evaporated on top of the structure). To satisfy the best conditions of the harmonics generation one should use structures with a small grating period $a$ and a low electron density $n_s$. For example, the maximum of the second harmonic is expected at incident wave frequency $\simeq 1$ THz (the generated second harmonic at $\simeq 2$ THz) in a structure with $n_s\simeq 4\times 10^{10}$/cm$^2$ and $a\simeq 0.14$ $\mu$m. In the second harmonic experiments one should use an asymmetric grating (see, e.g., \cite{Wang12}) to violate the central symmetry of the system (alternatively one can use the attenuated total reflection technique). The third-harmonic effect can be observed with a standard (symmetric) grating.   

To summarize, we have derived, within the self-consistent nonlinear response theory, exact analytical expressions for the second- and third-order polarizabilities of the $d$-dimensional uniform electron gas, and the corresponding (second-order) self-consistent response function. We have determined the optimal conditions for the second- and third-harmonic generation in low-dimensional electron systems and proposed experiments which could be used for the creation of terahertz frequency multipliers.

The financial support of this work by the Deutsche Forschungsgemeinschaft is gratefully acknowledged.

\bibliography{../../BIB-FILES/zerores,../../BIB-FILES/thz,../../BIB-FILES/emp,../../BIB-FILES/lowD,../../BIB-FILES/mikhailov,../../BIB-FILES/math,../../BIB-FILES/graphene}
\bibliographystyle{apsrev}
\end{document}